\documentclass[conference]{IEEEtran}
\IEEEoverridecommandlockouts

\usepackage[utf8]{inputenc}
\usepackage[T1]{fontenc}
\usepackage{tabularx}
\usepackage{multirow}
\usepackage{todonotes}
\usepackage{longtable}
\usepackage{lscape}
\usepackage{enumitem}
\usepackage{rotating}
\usepackage{microtype}
\usepackage{graphicx}
\usepackage{subfigure}
\usepackage{amssymb}
\usepackage{mathtools}
\usepackage{pifont}
\usepackage[final]{listings} 
\usepackage{color}
\usepackage{algorithm2e}
\usepackage{soulutf8}
\usepackage{epigraph}
\usepackage{tikz}
\usepackage[skip=2pt]{caption}

\usepackage{cite}
\usepackage{amsmath,amssymb,amsfonts}
\usepackage{algorithmic}
\usepackage{graphicx}
\usepackage{textcomp}
\usepackage[english]{babel}

\usepackage{url}

\newcommand{\hairspace}{\hspace{1pt}}
\newcommand{\eg}{\mbox{e.\hairspace{}g}.\ }  
\newcommand{\ie}{\mbox{i.\hairspace{}e}.\ }  

\usepackage{amssymb}

\usepackage{tikz}
\usetikzlibrary{shapes}
\usetikzlibrary{calc}
\tikzset{myptr/.style={-{Latex[scale=1.5]}}}

\tikzset{MyRoundedBox/.style={
		draw,
		rounded corners=3pt,
		inner sep=5pt,
		align=center
	}
}

\usepackage{listings}
\usepackage{cite}

\usepackage[binary-units=true]{siunitx}

\mathchardef\mhyphen="2D 

\usepackage{booktabs}
\usepackage{array}
\usepackage{multirow}

\usepackage{pifont}
\usepackage{expdlist}

\usepackage{fancyvrb}

\begin{document}
%

\title{A Performance and Resource Consumption Assessment of  Secure Multiparty Computation
\thanks{
This work has been supported by the German Federal Ministry of Education
and Research, project DecADe, grant 16KIS0538.
}
}

\author{\IEEEauthorblockN{ Marcel von Maltitz and Georg Carle}
\IEEEauthorblockA{ Technische Universität München, Department of Informatics \\ Chair for Network Architectures and Services
\\
85748 Garching b. München, Germany \\
\{vonmaltitz | carle\}@net.in.tum.de}

}


\maketitle

%

\newcommand{\FRESCO}{\textsc{Fresco}\xspace}
\newif\ifincludestackmem
\includestackmemfalse
\newif\iflongversion
\longversiontrue
\longversionfalse
\newcommand{\mytodo}[1]{\todo[inline]{#1}}

\newcommand{\numpeeroffsetcyclestwo}{5.16}
\newcommand{\numpeercoeffcyclestwo}{5.83556}
\newcommand{\numpeererrcyclestwo}{2.9451}

\newcommand{\numpeeroffsetcyclesnine}{15.263}
\newcommand{\numpeercoeffcyclesnine}{0.69823}
\newcommand{\numpeererrcyclesnine}{1.74056}

\newcommand{\numpeeroffsetpackets}{-2.743}
\newcommand{\numpeercoeffpackets}{2.69419}
\newcommand{\numpeererrpackets}{0.03332}

\newcommand{\numpeeroffsettime}{-1.086}
\newcommand{\numpeercoefftime}{2.01883}
\newcommand{\numpeererrtime}{0.24894}

\newcommand{\netlatoffsettime}{47.327}
\newcommand{\netlatcoefftime}{4.61851}
\newcommand{\netlaterrtime}{15415.50432}

\begin{abstract}
In recent years, secure multiparty computation (SMC) advanced from a theoretical technique to a practically applicable technology.
Several frameworks were proposed of which some are still actively developed.

We perform a first comprehensive study of performance characteristics of SMC protocols using a promising implementation based on secret sharing, a common and state-of-the-art foundation.
Therefor, we analyze its scalability with respect to environmental parameters as the number of peers, network properties -- namely transmission rate, packet loss, network latency -- and parallelization of computations as parameters and execution time, CPU cycles, memory consumption and amount of transmitted data as variables.

Our insights on the resource consumption show that such a solution is practically applicable in intranet environments and -- with limitations -- in Internet settings.

%

\end{abstract}
\IEEEpeerreviewmaketitle

\section{Introduction}
While the foundations for secure multiparty computation (SMC)
were laid about forty years ago \cite{Yao1982},  the topic experienced a revival in the last decade:
Starting as mere theoretic considerations, improvements in hardware performance made
practical implementations and productive use of SMC possible.
In consequence, a number of SMC frameworks emerged and its practical application was considered in research \cite{SugarBeetSMC,SEPIANetwork,SMCAviation,SMCFinancial}.

Their use cases have in common that they focus on singular events of orchestrated or manually triggered computations.
With Smart Buildings and the Internet of Things (IoT), a new type of use case for privacy-preserving data processing becomes relevant:
Regular and automated processing of data streams will be carried out on commodity or even low-end hardware.
Here, SMC can be the distributed system of choice for performing privacy-preserving aggregation of distributed data.
But this is only the case when the environmental constraints do not render its application infeasible:
Host nodes will only have a low amount of memory and a constrained CPU in terms of frequency and number of cores.
Furthermore, communication might happen via wireless LAN or even between different regions over the Internet.
Then, network performance is restricted in terms of limited transmission rate or high network latency and presence of packet loss.

It is hence vital to understand the resource requirements of a productively usable SMC solution. Our work provides these insights by performing a thorough performance evaluation  of a selected SMC framework based on secret sharing assessing the influence of a multitude of parameters on variables quantifying host (CPU and memory), network (transmitted data) and user resources (time) alongside with the identification of critical scaling behavior.

The remainder of the paper is structured as follows:
In Section \ref{sec:smc_overview}, we give an overview of SMC in general and argue for the framework we select for further examination.
Section \ref{sec:related_work} presents the related work regarding practical evaluation of SMC.
We present preliminary theoretical performance considerations for round based SMC protocols in Section \ref{sec:performance_model}.
Section \ref{sec:evaluation_setup} contains the description of our evaluation setup; the results are presented and discussed in Section \ref{sec:results}.
We elaborate the practical implications in Section \ref{sec:practical_implications}
and conclude our paper with Section \ref{sec:conclusion}.





\section{Secure Multiparty Computation}
\label{sec:smc_overview}
Secure multiparty computation enables multiple communicating parties to collaboratively compute a function
while being able to keep their respective input value completely confidential.
Yao initiated this field of research by presenting the Millionaire's Problem and the idea of Secure Function Evaluation \cite{Yao1982}\!\!\cite{Yao1986}.  
While many single purpose protocols were proposed, the main interest was in the creation of a general purpose framework which allows the computation of arbitrary functions.
Basic concepts were identified which allowed approaching this
\iflongversion
aim:
Using garbled circuits, functions are transformed in binary circuits, the truth table of every binary gate is obfuscated, so that the computing party is not able derive intermediary results. Homomorphic encryption allows computation on encrypted values and subsequent decryption of the final computation result.
Finally, secret sharing schemes with homomorphic properties allow to split private information among a group of participants and to compute predefined functions on the shares before recombining them to the collective final result.
\else
aim, most notably \emph{garbled circuits}\cite{Yao1986}, \emph{homomorphic encryption} \cite{Rivest1978a} and \emph{secret sharing schemes} \cite{Shamir1979}\cite{BGW88}.
\fi
Its theory flourished early in the 80's (cf.\@\cite{Goldreich1987,bmr,BGW88,Chaum1988,Chaum1988a,Rabin1989})
 while  implementations have only been developed in the last decade.  Among them, many have been proposed as proof of concept but were not publicly available \cite{SugarBeetSMC}
or have not been developed further since then \cite{SEPIANetwork} \cite{VIFF} \cite{FairplayMP}.
Currently, Sharemind \cite{Sharemind}, SPDZ-2 \cite{BristolMPC} \cite{MASCOT} and \FRESCO \cite{frescoURL}  constitute the state-of-the-art of actively developed SMC frameworks\footnote{There are further frameworks for the special two-party case, but they are not applicable in this multiparty context. }.

\paragraph{Sharemind}
Sharemind \cite{Sharemind} was mainly developed by Bodganov \cite{SharemindPhD}.
It is implemented in C++ and uses an additive secret sharing scheme in the ring \(\mathbb{Z}_{2^{32}}\).
Its focus was to provide a real-world suitable framework with appropriate performance.
They therefore opted to prevent only passive corruption -- which is less computationally expensive -- and to constrain their solution to strictly three computation parties, while allowing an arbitrary number of input parties.
Their argument is that further computation parties increase the communication overhead.
Sharemind is now part of the services that Cybernetica \cite{Cybernetica} provides. It is under active development but partially closed-source.

\paragraph{SPDZ-2}
SPDZ-2 \cite{BristolMPC} \cite{MASCOT} is currently developed by the University of Bristol.
The software is mainly written in C++ while the protocols can be written in Python.
It features the SPDZ protocol, which follows the current research direction of using additive secret sharing and performing a (computationally expensive) preprocessing phase in order to gain highly efficient protocol executions.

\paragraph{Framework for Efficient Secure Computation}
\FRESCO \cite{frescoURL} is developed by the Alexandra Institute in Denmark, a non-governmental organization for IT innovation and IT research.
It aims for being a non-prototypical, productively applicable generic SMC framework written in Java. I.e.\@ it is desired that the framework provides an abstraction from specific SMC primitives so that protocol specification can be performed independently. The benefits are that primitives can be switched afterwards while the specified protocol does not have to be changed. This especially enables simple incorporation of newest research results on SMC.

Currently, they support the Ben-Or--Goldwasser--Wigderson (BGW)~\cite{BGW88} protocol based on secret sharing using polynomials and the computation (``online'') phase of SPDZ \cite{BristolMPC} which has been successfully applied in \cite{Damgard2017a}. A full support of SPDZ is work in progress.

%

All of these solutions are secret sharing based. Hence, a similar performance behavior depending on the investigated parameters can be expected.
However, for our use case we need a solution which is able to
support computations with a theoretically arbitrary number of participants. This is not given by Sharemind.
Furthermore, Sharemind is closed-source which further obstructs assessment. SPDZ-2 is currently still work in progress on a level of fundamental changes and consequently not ready for a thorough performance measurement.
Our choice is therefore \FRESCO, which aims for production-ready application.

\section{Related Work}
\label{sec:related_work}
The newly gained interest in SMC during the last years resulted in a multitude of publications, which propose successive improvements or applications of established approaches\iflongversion{} (garbled circuits, linear secret sharing, homomorphic encryption, \ldots)\fi.
By contrast, the body of research is missing thorough performance measurements of SMC solutions.

Most of these publications do not provide performance data or merely a single result for their exact setting of application \cite{SugarBeetSMC} \cite{SMCFinancial} \cite{Thoma2012} \cite{Bogetoft2006} .
Others typically only evaluate overall execution time and to some lower degree transmitted bytes measured while at most varying the number of parties and the amount of input data \cite{SEPIANetwork} \cite{SMCAviation} \cite{Sharemind} \cite{Bonawitz2017,practicaltwoparty,hpsmc,comparisonperformance,smccommunicationcomplexity}.
Only few include further parameters like the transmission rate \cite{MASCOT} and technology-dependent factors like circuit size and depth \cite{FairplayMP} and evaluate further parameters e.g. throughput.

%
\vspace{2mm}
\noindent

For understanding whether SMC is also feasible in distributed systems, fog computing and the Internet of Things, it is necessary to perform more thorough measurements including further factors. It is vital to understand the influences of the network characteristics and to further examine the impact on host resources, i.e. CPU utilization and memory consumption.

We aim to provide the necessary insights by assessing the parameters
\emph{number of peers},
\emph{transmission rate},
\emph{network latency},
\emph{packet loss},
and \emph{input data parallelization}
while measuring the variables
\emph{execution time},
\emph{CPU cycles},
\emph{heap memory consumption}, and
\emph{transmitted bytes}.
%
\section{Preliminary Execution Time Considerations}
\label{sec:performance_model}
A possible theoretic computation model for secret sharing based SMC protocol foundations like BGW  ``is a complete synchronous network of $n$ processors''\cite{BGW88}.
The protocol itself, common to all processors, is dissected into rounds.
``In one round of computation each of the players can do an arbitrary amount of local computation, send a message to each of the players, and read all messages that were sent to it at this round''~\cite{BGW88}.
In secret sharing based protocols, such a message typically contains a share of a private local value --  \eg a polynomial in the BGW protocol --   held by the sender.
\newcommand{\commstepconsideration}{ We consider recombining the shares to be the last step $comp_m$. Hence, there are only $m-1$ communication steps. }

Considering the aforementioned rounds as a time factor,  the protocol becomes an alternating sequence\iflongversion\else\footnote{\commstepconsideration}\fi{} of local computation and network communication:
\begin{eqnarray}
  \label{eqn:sequence}
comp_1, comm_1,  \dots, comp_{m-1}, comm_{m-1}, comp_m
\end{eqnarray}
\iflongversion
\commstepconsideration
\fi

Furthermore, the communication steps are synchronization points for the
\iflongversion
players.
I.\,e., no player $P_i$ can already perform a computation $comp_{k+2}$, while another player $P_j$ still computes $comp_{k}$. Being a single step ahead is however possible for a single player when it locally possesses a polynomial while all other players still wait for a share of it in order to proceed.
\else
players, as they typically need the shares of the other participants in order to proceed with the next round.
\fi
\newline
We denote the costs in terms of time for a step $comp_i$ as ${cost}_{comp_i}$.
The message sent from player $P_k$ sent to $P_l$ during $comm_i$ is referred to as  ${msg}_{i,k \rightarrow l}$.

Two phases are typically common to all SMC protocols:
During the \emph{input phase} the own private input is transformed into shares and distributed among the participants. This takes one round.
In the \emph{output phase} the shares of the computed result are exchanged among all participants, so that each is able to recombine them and to obtain the plaintext result.
In \FRESCO this also takes a round\footnote{Some solution perform a resharing in order to make the final shares independent from the shares obtained in the computation. This is, e.g., necessary when the shares should be reused to perform further calculation. Then,  another round becomes necessary during this phase.}.

Regarding the basic arithmetic operations BGW provides, the round complexity varies. Addition is ``free'' as it does not need any communication\iflongversion{} and can be performed completely locally\fi. Multiplication requires rerandomization of the polynomial and the reduction of its degree\cite{BGW88}. This involves a step of communication between the participants, and hence, requires a round.

Theoretically, the communication cost of the $i$th round ${cost}_{{comm}_i}$ depends on the number of messages sent during the round.
As every participant sends an individual share of a polynomial to every other participant during communication steps, the overall number of shares sent is $\mathcal{O}(n^2)$.
Furthermore, every participant $p_i$ typically contributes its own input $v_i$ for the computation.
Hence, when a single multiplication step is specified in the protocol, this means that the product of all input values should be computed: $\prod_{i = 1}^{n} v_i$.
In such a case, $n-1$ single multiplication rounds are necessary; consequently the costs for such an array multiplication are $\mathcal{O}(n^3)$.

These theoretical costs assume a sequential execution of each communication.
However, inspection of the \FRESCO code and the analysis of its behavior show that sending and receiving for every participant can happen in parallel\footnote{One exception is the initial input sharing phase. Here, sending of shares is only performed by a single host at a time.}: Sending is a non-blocking action for the computation layer which hands over the messages to be sent to the communication layer of \FRESCO. Receiving is actually blocking on the computation layer, however, the communication layer is nevertheless able to receive all available messages simultaneously. In other words, waiting times for receiving multiple shares are not strictly additive.

When a host has sent out every share and it has received all other participants' shares, the next computation step can be performed.
So, in spite of the aforementioned theoretical complexity, due to parallelization the overall communication cost per round mainly depends on the pair of hosts, where communication takes longest:
\begin{eqnarray}
  \label{eqn:maxdelay}
{cost}_{{comm}_i} = \max_{1 \le k,  l \le n} {cost}_{{msg}_{i,k \rightarrow l}}
\end{eqnarray}

\iflongversion One dependency to the number of participants remains:\fi
While every round is practically performed in constant time,
the number of rounds per array multiplication increases linearly.

A further approximative simplification of the communication costs can be made:
Communication between two peers is always identically structured and bears shares as content. We could verify this claim using the \FRESCO code, which specifically only sends instances of the single class which represent the shares. Hence, we can simplify that
\begin{eqnarray}
  \label{eqn:costequality}
  \forall i  \in \{1, \ldots, m-1\}: {cost}_{comm_i} = {cost}_{comm}
\end{eqnarray}


%



Note that Equation \ref{eqn:costequality} does not hold for computation
\iflongversion
steps.
The input phase as well as the randomization/resharing step (\eg during multiplication) encompasses the generation of a new polynomial and the computation of single elements of it (which become the shares).
The output phase mainly includes the calculation of the Lagrange interpolation for reconstructing final computation result from the obtained shares.
Every single addition and multiplication of shares includes a single addition and multiplication of large integers\footnote{Represented as \texttt{BigInteger}s in \FRESCO.}.
\else
steps, as each phase performs different tasks.
\fi

Combining Equations \ref{eqn:sequence} and \ref{eqn:costequality} the overall costs of time can be estimated by
\begin{eqnarray}
  \label{eqn:costestimate}
  {cost}_{overall} = \sum_{i = 1}^{m} {cost}_{comp_i} + (m-1) * {cost}_{comm}
\end{eqnarray}

Using the model of the alternating sequence, two types of influences on the duration become visible:
The computation performance depends in the properties of the participants, the communication performance depends on the properties of \iflongversion the network links between them\else their network links\fi.
Due to the synchronizing behavior of rounds, the costs of both sides add up to the overall costs. 

In the first part of our measurements (Section \ref{sec:host} to \ref{sec:user}), we focus on how the overall costs are influenced by an increasing number of participants as well as network parameters, namely network latency, transmission rate and packet loss. Besides duration measurements, we also assess the memory footprint, the CPU  utilization and the amount of data transferred.
In the second part (Section \ref{sec:paral}), we examine whether the described sequential processing and the resulting additivity of communication and computation costs can be circumvented. Therefor, we regard cases, where multiple individual computation sessions can be parallelized instead of being executed sequentially. We show that the amortized costs per session decrease as a consequence.

%
\subsection*{Performance Comparison}

Conceptually, SMC replaces a Trusted Third Party (TTP) \iflongversion for collaboratively computing a common function\fi
  by providing a secure protocol implementation.
  Canetti \cite{Canetti1999} used this understanding to propose a now well-established method \iflongversion{} -- the real/ideal paradigm --  \fi to prove secrecy and correctness of an SMC approach.
  \iflongversion
  The method is to show an isomorphism between the \emph{real world} where SMC is applied and an \emph{ideal world} where a TTP handles the computation.
  \fi

  We can also apply this understanding \iflongversion not only to prove secrecy and correctness, but \fi to assess the performance penalty that SMC introduces.
  The ideal world which uses a TTP for computation \iflongversion is an alternative solution for the problem SMC solves, hence, it \fi can also be used as a performance baseline.
  In fact, in today's productively used systems, TTP solutions are the established standard; hence, the comparison with a TTP is also practically relevant.

  In order to do so, we align the necessary actions when using a TTP with the phases of an SMC computation.
  In a TTP setting, the input phase can be understood as providing the input data to the TTP.
  The output phase, in turn, comprises sending the result from the TTP to the participants.
  Computation steps themselves can be directly adapted.
  The whole comparison applied to the BGW protocol is shown in Table  \ref{tab:comparison}.

  \iflongversion
  In greater level of detail a further vital distinction has to be made:
  Choosing SMC as solution does not only change the number of steps to be performed and the structure of interaction but also the processed data:
  In our use case, we initially handle input values with the type \texttt{double}.
  A single primitive value has the length of 8\,Bytes.
  On the other side, the primitive value of the BGW protocol suite is a \emph{share}.
  This consists of a $(x,y)$ tuple denoting a point on the local polynomial.
  In \FRESCO the $x$ is stored as a \texttt{byte} and $y$ as \texttt{BigInteger}.
  By inspection we could find out, that $y$ typically has a length of 8 -- 12\,Bytes.
  \fi

  Presenting our results in section \ref{sec:results},  we add -- where applicable and foreseeable -- an estimation how a TTP solution would perform\iflongversion{} in the shown experiments\fi.
  In these cases we approximate the communication performance as described before while neglecting the comparatively low influence of
  the computation steps.

  \begin{table*}
    \begin{center}
      \begin{tabular}{l*{2}l*{2}l}

        \toprule
        & \multicolumn{2}{c}{SMC}                                                          & \multicolumn{2}{c}{TTP} \\
        Phase          & Computation per host                                                             & Communication (overall) & Computation on TTP    & Communication (overall) \\
        \midrule
        Close          & Generation of polynomial, calculation of n shares                                  & $n^2 - n$ messages      & ---                   & n messages  \\
        Addition       & $n-1$ additions                                                                  & ---                     & $n-1$ additions       & --- \\
        Multiplication & $n-1$ multiplications, $\text{Comp}_\mathit{Close}$, $\text{Comp}_\mathit{Open}$ & $n^2 - n$ messages      & $n-1$ multiplications & --- \\
        Open           & Lagrange interpolation                                                           & $n^2 - n$ messages      & ---                   & n messages \\
        \bottomrule

      \end{tabular}
    \end{center}
    \caption{Performance comparison SMC vs. TTP}
  \vspace{-2mm}
    \label{tab:comparison}
  \end{table*}



\section{Evaluation Setup}
In the following we describe our test setup.
The use case explains which computations have been carried out via SMC.
Here, we refer to a real-world use case performed at our lab; the functionality is however similar to other real-world systems.
In the second part, the methodology, we document the measurement environment in terms of used software, hardware and measurement tools.
\label{sec:evaluation_setup}
\subsection{Use Case}
\label{sec:usecase}
Our use case is inspired by MeasrDroid~\cite{measrdroid}, a smartphone app which allows users to gain insights in the sensor data of their smartphones and allows comparison to other users.
We assume a set of moving devices\iflongversion{} which are aware of their own location using a GPS module\fi.
One property of interest is the summed and averaged travel distance over the set of devices.

Insecurely and without SMC, the functionality is realized as follows:
Each client is able to derive a stream of distances from the raw GPS coordinates\iflongversion{} by calculating the distance between each pair of successive coordinates\fi.
They can connect to a common central trusted server\iflongversion, which has the purpose to collect statistics about them\fi.
Upon each connection the client transmits the travel distance since its last connection.
These distances are collected as a running average\iflongversion{} (by storing the sum and the number of submissions separately)\fi.
At any given point in time the overall average distance can then be computed by the server\iflongversion{} by dividing the stored distance sum by the number of submissions\fi.

%
%
%
%
%
%

%
%
%

%

In order to apply \FRESCO\iflongversion{} to this problem\fi, the input has to be organized in synchronous sessions.
In every session, each device contributes its distance since the last session\iflongversion\footnote{The device can either measure its GPS position at the beginning of each session or poll its location at an arbitrary frequency and sum up the determined distance. The latter approach is more exact than the former.}\fi, whereas the statistics server inputs the current value of the running sum (starting with 0).
\iflongversion By doing so, the server fulfills the technical requirement of also contributing a value while it does not semantically change the computed result.\fi
The result of each round is \iflongversion in turn \fi saved by the statistics server.

\iflongversion
With regard to the privacy of the system, it is uncritical that the total sum is calculated privately, while the division step to retrieve the average is performed without SMC.
Under the premise that the number of participants is known (which is necessary to perform the sharing correctly) the total sum would always be derivable from the average value.
Consequently, the total sum does not leak any more information than the average does.
\fi
Knowing that communication between the peers is the typical bottleneck for SMC \cite{SMCAviation} \cite{SMCFinancial} \cite{Bonawitz2017}, the choice of the use case is beneficial for our performance measurement:
The computational part is comparatively low so that performance effects caused by communication and their relationship to the named parameters become clearly visible. This allows better assessment of the communication bottleneck of SMC based solutions with negligible influence by the local computations.

\textit{Input Data}:
We used real world data retrieved from MeasrDroid, yielding five traces consisting of 20000\,GPS tuples each, being collected in intervals of 15 to 20 minutes depending on the individual configuration per device.
The utilization of more nodes for some measurements  made it necessary to provide more input data to be used for the computation.
The data itself does in no way influence the performance of the system. Hence, without loss of application and closeness to reality we took our data from the original 5 donors and duplicated the inputs until every used node had an own list of GPS input tuples.

\iflongversion
In order to evaluate the framework we wrote a lightweight wrapper which loads input data from a local storage, performs necessary  preprocessing and starts the computation using \FRESCO.
\fi


\subsection{Methodology}
We evaluated \FRESCO in the following setting:
\subsubsection{Hardware and Host Setup}
\label{sec:hardware_setup}
For our tests we had 15 physical hosts available.
Each host has an Intel Xeon \iflongversion E3-1265L V2 \fi CPU with eight cores at 2.50\,GHz and a cache size of 8192\,KB. They have 15.780\,MB of RAM each and a 1\,Gbit networking interface.
They are arranged in the shape of a star topology, all hosts are connected via a single switch.
The default link latency is around 0.18\,ms and there is no packet loss.
The test hosts \iflongversion use a dedicated PXE server to boot from. This server provides an image of Debian Jessie (8.5) using a 3.16 Linux kernel.
\else
use Debian Jessie (8.5) and a 3.16 Linux kernel.
\fi
The source code is compiled to a Java application which is in turn executed by the Java VM from the OpenJDK 1.8.0\_111.
For some tests simulating intranet, Internet and mobile Internet settings we added an artificial delay of 16\,ms, 50\,ms, 200\,ms and 500\,ms to the communication round-trip
equally distributing the delay to both hosts of each link using \texttt{tc} \iflongversion\cite{lartc}\fi. For this purpose, \texttt{tc} delays every outgoing packet by the half of the desired additional delay. Packet loss is also simulated via \texttt{tc}.


\subsubsection{Software Setup}
We use an orchestration layer which configures all client-side parameters (latency, transmission rate, packet loss,  \dots) and then starts the target application.
\iflongversion
This is performed in an sequentially interleaved fashion, so that the state of the participating peers is held roughly synchronized.
\fi
The Java application itself locally loads GPS coordinates in order to perform 1000 executions of the protocol per measurement. This repetition makes the measurement results more robust against random performance fluctuations during single computations.
Furthermore, each measurement itself has been repeated 50 times if not noted otherwise.

\subsubsection{Measurement Software}
\iflongversion
Profiling is performed using \emph{perf} from the linux-tools (version 3.16+63), \emph{BTrace} \iflongversion\cite{btraceURL}\fi (version 1.3.8.3 (20160926)) and \emph{tshark} (version 2.2.4).

We use perf to count CPU cycles consumed by the executed process under test.
In our context, only the overall number of cycles is considered relevant.

BTrace is used to evaluate the memory consumption and the execution time of the application under test.
It performs profiling by bytecode tracing, \ie instrumenting the bytecode under test.
BTrace allows event-based and polling-based profiling. Both options are necessary in our context.
The profiling starts with execution of the Java process under test and stops with its termination; time counting is started when the specified methods are entered and stopped when they are left.
However, the memory consumption of the Java process changes during execution time, hence it must be polled frequently in order to gain insights on the behavior of the process.
In the measurements we detected when memory polling significantly influenced e.g. CPU measurements.
We then split the measurement of both variables in separate runs in order to avoid interferences.

We use tshark from the wireshark application to capture the TCP packet stream constituting the computation communication.
By collecting the whole data stream, it is ensured that analysis is not constrained by a priori assumptions during data collection, but the full potential of the measured information can be exploited.
\else
Profiling is performed using \emph{perf} from the linux-tools (version 3.16+63) for counting CPU cycles, \emph{BTrace} \iflongversion\cite{btraceURL}\fi (version 1.3.8.3 (20160926)) for assessing memory consumption and execution time and \emph{tshark} (version 2.2.4) from wireshark for collecting the raw transmitted data.
\fi

\section{Results}
\newcommand{\defaultcaption}[2]{Impact of #1 on #2}
\newcommand{\extime}{the execution time}
\newcommand{\packets}{the transmitted bytes}
\newcommand{\cpucycles}{the CPU cycles}
\newcommand{\cpuinst}{the number of CPU instructions}
\newcommand{\heapmem}{the maximum allocated heap memory}
\newcommand{\stackmem}{the maximum allocated stack memory}
\newcommand{\paralgaintable}{RDATA}
\newcommand{\numpeeroffset}{RDATA}
\newcommand{\numpeercoeff}{RDATA}
\newcommand{\numpointoffsetthree}{RDATA}
\newcommand{\numpointcoeffthree}{RDATA}
\newcommand{\numpointoffsetnine}{RDATA}
\newcommand{\numpointcoeffnine}{RDATA}
\newcommand{\numpointoffsetfifteen}{RDATA}
\newcommand{\numpointcoefffifteen}{RDATA}
\newcommand{\numpointlevela}{RDATA}
\newcommand{\numpointlevelb}{RDATA}
\newcommand{\numpointlevelc}{RDATA}
\newcommand{\paralgainone}{RDATA}
\newcommand{\paralgainfour}{RDATA}
\newcommand{\paralgainfive}{RDATA}
\newcommand{\netlatoffset}{RDATA}
\newcommand{\netlatcoeff}{RDATA}
\newcommand{\maxheapusage}{RDATA}

\newcommand{\bandwidthmbamountonembit}{RDATA}
\newcommand{\bandwidthmbamounttenmbit}{RDATA}
\newcommand{\bandwidthmbamounthundmbit}{RDATA}
\newcommand{\bandwidthmbamountonegbit}{RDATA}
\newcommand{\bandwidthmbamountbigpacketsonembit}{RDATA}

\newcommand{\reason}{}
\label{sec:results}

In the next subsection we focus on the host resources \emph{heap memory} and consumed \emph{CPU cycles}. Afterwards we analyze the amount of \emph{transmitted data} representing the network resource.
Then the \emph{execution duration}, most directly affecting the user, is discussed. As the last subsection, we analyse the influence of parallelization of computations on the named resources.
\subsection{Host Resources}
\label{sec:host}

%
\subsubsection{RAM}
\label{sec:memory}
In our context, RAM is separated in stack and heap memory.
Our measurements showed that stack memory always ranged from 16\,MBytes to 20\,MBytes.
We consider these variations to be negligible. Therefore, we focus completely on
heap memory consumption in the following.

\message{Including measurement diagram: figures/smc/smc_max_memory_heap_by_nodes_paper}
\begin{figure}[t!]
\resizebox{\columnwidth}{!}{%
  \input{figures/smc/smc_max_memory_heap_by_nodes_paper.tikz}
}
\caption{\defaultcaption{the number of peers}{\heapmem}}
\label{fig:smc_max_memory_heap_by_nodes}
\end{figure}

Our baseline execution with 3 peers and a setup as described in Section~\ref{sec:hardware_setup}, the standard memory consumption is around 69\,MBytes.
This is only negligibly influenced by networking parameters. The delay in execution caused by  a lower transmission rate, higher network latency or packet loss
typically influences the speed of memory allocation and in consequence the garbage collector. This yields variations up to \(\pm 3\)\,MBytes.

We identified a strong correlation when scaling the number of peers.
Increasing this parameter, heap memory consumption gradually diverges step-wise (cf. Figure~\ref{fig:smc_max_memory_heap_by_nodes}).
The \FRESCO application uses around 70\,MBytes during a computation with 3 to 5 peers, which increments to
530\,MBytes for 7 to 9 peers and increases again to 840\,MBytes for 11 to 15 peers.
This is expectable as data about current connections as well as intermediate results like the shares of all other participants are stored on the heap.
We deduce a linear trend from Figure \ref{fig:smc_max_memory_heap_by_nodes} where the notable amount of outliers at x=15 already foreshadows the next step of heap increment.
In any case, this factor rapidly becomes critical:
With 15 peers, \FRESCO already starts exceeding the memory resources of a Raspberry Pi \cite{raspi} 3 B (1GByte RAM) and uses a considerate amount of the memory of a current smartphone, where typically 2\,GByte to 4\,GByte are available for the whole system and all concurrently running applications.
\subsubsection{CPU Cycles}

\message{Including measurement diagram: figures/smc/smc_cycles_by_netlat_no_paral_paper}
\begin{figure}[t!]
\resizebox{\columnwidth}{!}{%
  \input{figures/smc/smc_cycles_by_netlat_no_paral_paper.tikz}
}
\caption{\defaultcaption{network latency}{\cpucycles}}
\label{fig:smc_cycles_by_netlat}
\end{figure}

During the CPU measurements we noticed that there is a major difference in the number of consumed CPU cycles when comparing a fixed node (in our case, node 2) with the last node (having the highest ID) in the set of participants.
\label{sec:setup_phase}
This difference is not an effect of the actual computation, but the reason is rather found in the setup phase of \FRESCO.
The initial step before coordinating the computation, the hosts have to establish connections with every other participant.
This is achieved by every application listening for incoming connections and performing own connection attempts to other hosts in parallel, driven by busy waiting.

During our measurements, the application was started on all hosts with increasing ID, always having a little delay between the invocations.
Due to this reason, that phase exhibits a specific pattern:
The first application starts to poll for all other hosts which are not yet listening for incoming connections.
This requires a notable number of CPU cycles.
When the second application comes up, it immediately connects to the first host due to one of its connection attempts.
From this point in time, both hosts poll in order to connect to all other hosts.

In consequence, the first hosts performs most polling while waiting for not yet started participants, while the last host needs only a comparatively small amount of TCP SYN attempts before all other hosts connected to it, wasting much less CPU cycles.
This understanding is necessary to interpret our results.

Our baseline is around \(21.5 * 10^{9}\) cycles for the first peer node and \(16.5 * 10^{9}\) cycles for the last node.
When reducing network performance as described in Section~\ref{sec:memory}, consumption drops to approximately \(12.5 * 10^{9}\).
This effect ist best depicted in the cases of network latency (Figure \ref{fig:smc_cycles_by_netlat}) and packet loss (Figure \ref{fig:smc_cycles_by_packetloss}) and can be attributed to the previously explained startup phase: Impeded transmission adds another constrain on the polling which in turn becomes slower and less CPU intensive.

\message{Including measurement diagram: figures/smc/smc_cycles_by_packetloss_paper}
\begin{figure}[t!]
\resizebox{\columnwidth}{!}{%
  \input{figures/smc/smc_cycles_by_packetloss_paper.tikz}
}
\caption{\defaultcaption{packet loss}{\cpucycles}}
\label{fig:smc_cycles_by_packetloss}
\end{figure}

Additionally Figure \ref{fig:smc_cycles_by_netlat} shows a slight increase in CPU cycles when increasing the network latency further.
As the number of instructions did not increase during the same measurements, we expect this effect to be caused by IO waiting time during the delayed protocol execution.

%

On the side of number of participating nodes, the number of consumed CPU cycles depends strictly linear on it.
For the first node we get (MSE\footnote{Mean squared error}: \numpeererrcyclestwo)
\[(\numpeeroffsetcyclestwo + \numpeercoeffcyclestwo * n) * 10^9 \]
and for the last node (MSE: \numpeererrcyclesnine)
\[ (\numpeeroffsetcyclesnine + \numpeercoeffcyclesnine * n) * 10^9 \]
We see that the amount of CPU cycles used in the startup phase heavily outweighs the increase of participating nodes.

\subsection{Network Resources}
\label{sec:net}
%

Our baseline of transmitted data for three peers is 5.35\,MBytes per peer.

We identified that the amount of transmitted data per peer varies around 400\,KBytes upon network changes.
By package inspection a common reason could be found in the network communication behavior of \FRESCO:

\label{sec:packet_explanation}
The communication layer of \FRESCO on the host of sender $s$ receives and buffers a serialized object $o_{r,1}$ from the computation layer to be sent to recipient $r$.
The actual transmission of $o_{r,1}$ happens in the moment when $r$ is prepared to accept the data.
However, this action does not block on the sender side. I.e. if the sender itself does not have to wait for any further incoming data from other peers, it can proceed with the next computation step immediately.
Here, it can already prepare the next step of communication, including to create and prepare the next object $o_{r,2}$ to be sent to the same recipient.
Given the recipient did not request $o_{r,1}$ at this point in time, the communication layer will combine $o_{r,1}$ and $o_{r,2}$ into a single message, which is then sent to $r$ when its request happens.
Combination of packets results in a reduction of the overall amount to be sent by reducing the absolute number of necessary  packet headers.
It is coincidence that this effect is most useful in environments with constrained transmission, where it also naturally happens most often.

The measurements of two network parameters reflect this behavior up to some degree:
When reducing the transmission rate to 1\,MBit, a drop to 5.10\,MBytes can be detected.
A similar behavior occurs when adding artificial network latency, however, without a distinct trend.

\message{Including measurement diagram: figures/smc/smc_bytes_by_packetloss_paper}
\begin{figure}[t!]
\resizebox{\columnwidth}{!}{%
  \input{figures/smc/smc_bytes_by_packetloss_paper.tikz}
}
\caption{\defaultcaption{packet loss}{the transmitted KBytes}}
\label{fig:smc_bytes_by_packetloss}
\end{figure}

While these deviations undercut the baseline,  packet loss yields an increase of the amount of transmitted data (cf.  Figure~\ref{fig:smc_bytes_by_packetloss}).
This behavior is expected as packet loss requires retransmissions. With a maximum of 10\,\% packet loss, transmitted data was increased by approximately 400\,KBytes.

Regarding the number of peers, the number of messages to be exchanged between all peers depends quadratically on it.
Our measurements support this by showing that the amount of transferred bytes between \emph{a pair of hosts} increases linearly.
In our setting the increase follows the following regression line (MSE: \numpeererrpackets):
\[(\numpeeroffsetpackets + \numpeercoeffpackets * n) \text{ MBytes}\]
In other words, for each peer approximately 2\,MByte of additional data is transmitted \emph{per host}.

\subsection{User Resource: Time}
\label{sec:user}
The computation duration is the most interesting variable from the user perspective.
Our baseline is 5.35 seconds for 1000 calculations, \ie each computation costs around 5\,ms, whereas the startup of the Java Virtual Machine is not included.

We can see that time is heavily and differently influenced by the evaluated parameters:
The increase in time is strictly linear when adding more participants.
At first, this might surprise as the exchanged messages between all participants increase quadratically in their number. However, in Section~\ref{sec:performance_model} we already elaborated how parallel execution of communication can reduce the complexity by \(n\).
As a regression function (cf. Figure~\ref{fig:smc_executiontime_by_peers}) we yield (MSE: \numpeererrtime):
\[(\numpeeroffsettime + \numpeercoefftime * n) \text{ ms}\]

As comparison, the communication delay of a TTP solution does not notably depend on the number of participants, given that sending and receiving messages can happen in parallel.

Network latency also causes a linear increase. This directly corresponds with the intuition
that every message is delayed by a constant factor. However, the influence is notably stronger in absolute terms.
The following regression function (cf. Figure~\ref{fig:smc_executiontime_by_netlat_no_paral}) holds for three participating peers (MSE: \netlaterrtime):
\[\netlatoffsettime ms + \netlatcoefftime * \textit{network latency}\]

Execution inside an intranet takes around 4 seconds for 1000 sequential computations.
When communicating via the Internet (50\,ms to 300\,ms), the computations already cost 5 to 25 minutes.
The magnitude of the duration can be roughly estimated as follows:
During the input phase with $n = 3 $ participating hosts, $n * (n-1) = 6 $ messages have to be exchanged.
Each participant sequentially waits for $ n-1 = 2 $ messages from the other parties.
The performed addition operation is free of communication.
During the output phase, again 6 messages have to be exchanged, but this time waiting is performed in parallel\footnote{Using Equation \ref{eqn:maxdelay} we count this as a single message.}.
Hence, as an estimate in our setup, every participant sequentially waits for $n = 3$ messages, which can consist of one to two packets each.
A message of one packet costs a single network delay. A message of two packets costs three times the network delay as the second packet is only sent when the sender has received an acknowledgement message from the recipient.
In consequence, we gain an interval of $[n * \text{network latency}, 3n * \text{network latency}]$ per protocol execution. 

\message{Including measurement diagram: figures/smc/smc_executiontime_by_nodes_paper}
\begin{figure}[t!]
\resizebox{\columnwidth}{!}{%
\begin{tikzpicture}[x=1pt,y=1pt]
\definecolor{fillColor}{RGB}{255,255,255}
\path[use as bounding box,fill=fillColor,fill opacity=0.00] (0,0) rectangle (252.94, 93.95);
\begin{scope}
\path[clip] (  0.00,  0.00) rectangle (252.94, 93.95);
\definecolor{drawColor}{RGB}{255,255,255}
\definecolor{fillColor}{RGB}{255,255,255}

\path[draw=drawColor,line width= 0.6pt,line join=round,line cap=round,fill=fillColor] ( -0.00,  0.00) rectangle (252.95, 93.95);
\end{scope}
\begin{scope}
\path[clip] ( 24.86, 22.87) rectangle (188.58, 93.95);
\definecolor{fillColor}{RGB}{255,255,255}

\path[fill=fillColor] ( 24.86, 22.87) rectangle (188.58, 93.95);
\definecolor{drawColor}{gray}{0.92}

\path[draw=drawColor,line width= 0.3pt,line join=round] ( 24.86, 35.31) --
	(188.58, 35.31);

\path[draw=drawColor,line width= 0.3pt,line join=round] ( 24.86, 57.97) --
	(188.58, 57.97);

\path[draw=drawColor,line width= 0.3pt,line join=round] ( 24.86, 80.62) --
	(188.58, 80.62);

\path[draw=drawColor,line width= 0.6pt,line join=round] ( 24.86, 23.99) --
	(188.58, 23.99);

\path[draw=drawColor,line width= 0.6pt,line join=round] ( 24.86, 46.64) --
	(188.58, 46.64);

\path[draw=drawColor,line width= 0.6pt,line join=round] ( 24.86, 69.29) --
	(188.58, 69.29);

\path[draw=drawColor,line width= 0.6pt,line join=round] ( 24.86, 91.95) --
	(188.58, 91.95);

\path[draw=drawColor,line width= 0.6pt,line join=round] ( 38.50, 22.87) --
	( 38.50, 93.95);

\path[draw=drawColor,line width= 0.6pt,line join=round] ( 61.24, 22.87) --
	( 61.24, 93.95);

\path[draw=drawColor,line width= 0.6pt,line join=round] ( 83.98, 22.87) --
	( 83.98, 93.95);

\path[draw=drawColor,line width= 0.6pt,line join=round] (106.72, 22.87) --
	(106.72, 93.95);

\path[draw=drawColor,line width= 0.6pt,line join=round] (129.46, 22.87) --
	(129.46, 93.95);

\path[draw=drawColor,line width= 0.6pt,line join=round] (152.19, 22.87) --
	(152.19, 93.95);

\path[draw=drawColor,line width= 0.6pt,line join=round] (174.93, 22.87) --
	(174.93, 93.95);
\definecolor{fillColor}{RGB}{0,0,0}

\path[fill=fillColor] ( 38.50, 35.96) circle (  1.96);

\path[fill=fillColor] ( 61.24, 44.15) circle (  1.96);

\path[fill=fillColor] ( 83.98, 53.06) circle (  1.96);

\path[fill=fillColor] (106.72, 62.45) circle (  1.96);

\path[fill=fillColor] (129.46, 71.70) circle (  1.96);

\path[fill=fillColor] (152.19, 80.75) circle (  1.96);

\path[fill=fillColor] (174.93, 90.72) circle (  1.96);
\definecolor{drawColor}{RGB}{0,0,0}

\path[draw=drawColor,line width= 1.1pt,line join=round] ( 38.50, 35.24) --
	( 61.24, 44.39) --
	( 83.98, 53.54) --
	(106.72, 62.68) --
	(129.46, 71.83) --
	(152.19, 80.98) --
	(174.93, 90.13);
\definecolor{drawColor}{gray}{0.40}

\path[draw=drawColor,line width= 0.6pt,line join=round] ( 38.50, 26.10) --
	( 61.24, 26.10) --
	( 83.98, 26.10) --
	(106.72, 26.10) --
	(129.46, 26.10) --
	(152.19, 26.10) --
	(174.93, 26.10);
\definecolor{drawColor}{gray}{0.20}

\path[draw=drawColor,line width= 0.6pt,line join=round,line cap=round] ( 24.86, 22.87) rectangle (188.58, 93.95);
\end{scope}
\begin{scope}
\path[clip] (  0.00,  0.00) rectangle (252.94, 93.95);
\definecolor{drawColor}{gray}{0.30}

\node[text=drawColor,anchor=base east,inner sep=0pt, outer sep=0pt, scale=  0.64] at ( 19.91, 21.78) {0};

\node[text=drawColor,anchor=base east,inner sep=0pt, outer sep=0pt, scale=  0.64] at ( 19.91, 44.44) {10};

\node[text=drawColor,anchor=base east,inner sep=0pt, outer sep=0pt, scale=  0.64] at ( 19.91, 67.09) {20};

\node[text=drawColor,anchor=base east,inner sep=0pt, outer sep=0pt, scale=  0.64] at ( 19.91, 89.74) {30};
\end{scope}
\begin{scope}
\path[clip] (  0.00,  0.00) rectangle (252.94, 93.95);
\definecolor{drawColor}{gray}{0.20}

\path[draw=drawColor,line width= 0.6pt,line join=round] ( 22.11, 23.99) --
	( 24.86, 23.99);

\path[draw=drawColor,line width= 0.6pt,line join=round] ( 22.11, 46.64) --
	( 24.86, 46.64);

\path[draw=drawColor,line width= 0.6pt,line join=round] ( 22.11, 69.29) --
	( 24.86, 69.29);

\path[draw=drawColor,line width= 0.6pt,line join=round] ( 22.11, 91.95) --
	( 24.86, 91.95);
\end{scope}
\begin{scope}
\path[clip] (  0.00,  0.00) rectangle (252.94, 93.95);
\definecolor{drawColor}{gray}{0.20}

\path[draw=drawColor,line width= 0.6pt,line join=round] ( 38.50, 20.12) --
	( 38.50, 22.87);

\path[draw=drawColor,line width= 0.6pt,line join=round] ( 61.24, 20.12) --
	( 61.24, 22.87);

\path[draw=drawColor,line width= 0.6pt,line join=round] ( 83.98, 20.12) --
	( 83.98, 22.87);

\path[draw=drawColor,line width= 0.6pt,line join=round] (106.72, 20.12) --
	(106.72, 22.87);

\path[draw=drawColor,line width= 0.6pt,line join=round] (129.46, 20.12) --
	(129.46, 22.87);

\path[draw=drawColor,line width= 0.6pt,line join=round] (152.19, 20.12) --
	(152.19, 22.87);

\path[draw=drawColor,line width= 0.6pt,line join=round] (174.93, 20.12) --
	(174.93, 22.87);
\end{scope}
\begin{scope}
\path[clip] (  0.00,  0.00) rectangle (252.94, 93.95);
\definecolor{drawColor}{gray}{0.30}

\node[text=drawColor,anchor=base,inner sep=0pt, outer sep=0pt, scale=  0.64] at ( 38.50, 13.51) {3};

\node[text=drawColor,anchor=base,inner sep=0pt, outer sep=0pt, scale=  0.64] at ( 61.24, 13.51) {5};

\node[text=drawColor,anchor=base,inner sep=0pt, outer sep=0pt, scale=  0.64] at ( 83.98, 13.51) {7};

\node[text=drawColor,anchor=base,inner sep=0pt, outer sep=0pt, scale=  0.64] at (106.72, 13.51) {9};

\node[text=drawColor,anchor=base,inner sep=0pt, outer sep=0pt, scale=  0.64] at (129.46, 13.51) {11};

\node[text=drawColor,anchor=base,inner sep=0pt, outer sep=0pt, scale=  0.64] at (152.19, 13.51) {13};

\node[text=drawColor,anchor=base,inner sep=0pt, outer sep=0pt, scale=  0.64] at (174.93, 13.51) {15};
\end{scope}
\begin{scope}
\path[clip] (  0.00,  0.00) rectangle (252.94, 93.95);
\definecolor{drawColor}{RGB}{0,0,0}

\node[text=drawColor,anchor=base,inner sep=0pt, outer sep=0pt, scale=  0.80] at (106.72,  2.50) {Number of Peers [\#]};
\end{scope}
\begin{scope}
\path[clip] (  0.00,  0.00) rectangle (252.94, 93.95);
\definecolor{drawColor}{RGB}{0,0,0}

\node[text=drawColor,rotate= 90.00,anchor=base,inner sep=0pt, outer sep=0pt, scale=  0.80] at (  5.51, 58.41) {Time [s]};
\end{scope}
\begin{scope}
\path[clip] (  0.00,  0.00) rectangle (252.94, 93.95);
\definecolor{fillColor}{RGB}{255,255,255}

\path[fill=fillColor] (199.96, 34.39) rectangle (252.95, 82.42);
\end{scope}
\begin{scope}
\path[clip] (  0.00,  0.00) rectangle (252.94, 93.95);
\definecolor{drawColor}{RGB}{0,0,0}

\node[text=drawColor,anchor=base west,inner sep=0pt, outer sep=0pt, scale=  0.80] at (199.96, 71.91) {Execution type};
\end{scope}
\begin{scope}
\path[clip] (  0.00,  0.00) rectangle (252.94, 93.95);
\definecolor{fillColor}{RGB}{255,255,255}

\path[fill=fillColor] (199.96, 53.85) rectangle (214.41, 68.30);
\end{scope}
\begin{scope}
\path[clip] (  0.00,  0.00) rectangle (252.94, 93.95);
\definecolor{fillColor}{RGB}{0,0,0}

\path[fill=fillColor] (207.18, 61.07) circle (  1.96);
\end{scope}
\begin{scope}
\path[clip] (  0.00,  0.00) rectangle (252.94, 93.95);
\definecolor{drawColor}{RGB}{0,0,0}

\path[draw=drawColor,line width= 1.1pt,line join=round] (201.40, 61.07) -- (212.97, 61.07);
\end{scope}
\begin{scope}
\path[clip] (  0.00,  0.00) rectangle (252.94, 93.95);
\definecolor{drawColor}{RGB}{0,0,0}

\path[draw=drawColor,line width= 0.6pt,line join=round] (201.40, 61.07) -- (212.97, 61.07);
\end{scope}
\begin{scope}
\path[clip] (  0.00,  0.00) rectangle (252.94, 93.95);
\definecolor{fillColor}{RGB}{255,255,255}

\path[fill=fillColor] (199.96, 39.39) rectangle (214.41, 53.85);
\end{scope}
\begin{scope}
\path[clip] (  0.00,  0.00) rectangle (252.94, 93.95);
\definecolor{fillColor}{gray}{0.40}

\path[fill=fillColor] (207.18, 49.67) --
	(209.83, 45.09) --
	(204.54, 45.09) --
	cycle;
\end{scope}
\begin{scope}
\path[clip] (  0.00,  0.00) rectangle (252.94, 93.95);
\definecolor{drawColor}{gray}{0.40}

\path[draw=drawColor,line width= 1.1pt,line join=round] (201.40, 46.62) -- (212.97, 46.62);
\end{scope}
\begin{scope}
\path[clip] (  0.00,  0.00) rectangle (252.94, 93.95);
\definecolor{drawColor}{gray}{0.40}

\path[draw=drawColor,line width= 0.6pt,line join=round] (201.40, 46.62) -- (212.97, 46.62);
\end{scope}
\begin{scope}
\path[clip] (  0.00,  0.00) rectangle (252.94, 93.95);
\definecolor{drawColor}{RGB}{0,0,0}

\node[text=drawColor,anchor=base west,inner sep=0pt, outer sep=0pt, scale=  0.64] at (216.22, 58.87) {SMC};
\end{scope}
\begin{scope}
\path[clip] (  0.00,  0.00) rectangle (252.94, 93.95);
\definecolor{drawColor}{RGB}{0,0,0}

\node[text=drawColor,anchor=base west,inner sep=0pt, outer sep=0pt, scale=  0.64] at (216.22, 44.42) {TTP (est.)};
\end{scope}
\end{tikzpicture}
}
\caption{\defaultcaption{number of peers}{\extime}}
\label{fig:smc_executiontime_by_peers}
\end{figure}

\message{Including measurement diagram: figures/smc/smc_executiontime_by_netlat_no_paral_paper}
\begin{figure}[t!]
\resizebox{\columnwidth}{!}{%
  \input{figures/smc/smc_executiontime_by_netlat_no_paral_paper.tikz}
}
\caption{\defaultcaption{network latency}{\extime}}
\label{fig:smc_executiontime_by_netlat_no_paral}
\end{figure}

Utilizing a basic TTP solution, all hosts send their data during a single network delay. The computation itself is performed locally. At the end another network delay is added for  sending the results to all participants (in parallel).
While it seems that the performance of the SMC solution is acceptably worse in comparison, it is important to note that the duration of the SMC session scales proportionally with increasing number of peers, while the TTP does not depend on this factor (cf. Figure~\ref{fig:smc_executiontime_by_peers}).

When packet loss occurs, repeated retransmissions become necessary.  
Due to this, we expect the execution time (Figure~\ref{fig:smc_executiontime_by_packetloss})\iflongversion\footnote{Note: Starting at a packet loss rate of 6.0\,\% we reduce the amount of repetitions per test case due to the increased length of the measurements.}\fi{} to constitute a geometric
\iflongversion
row:
Having a fixed percentage \(p_{loss}\) of packet loss,  \(p_{loss} * |packets|_{ p_{loss} == 0}\) have to be repeated.
From this part, the percentage \(p_{loss}\) has to be repeated again, as it got lost during retransmission. This continues until all packets have been sent.

Effectively, this constitutes the following row,
row. This also has an analytical equivalent as \(p_{loss} < 1\):

\begin{eqnarray}
\sum^\infty_{k = 0} p_{loss}^k  = \frac{1}{1-p_{loss}}
\end{eqnarray}

This function increases
\else
row and to increase
\fi
hyperbolically in the interval \([0, 1[\) with increasing packet loss probability \(p_{loss}\).
One would expect the same characteristics for the execution time.
However, the steep increase only happens very late when \(p_{loss}\) is near 1. The analyzed interval from 0\,\% to 10\,\% is at the beginning of the function's domain, where only a linear increase becomes visible.
\iflongversion

We could not increase the packet loss further in order to show the hyperbolical increase as the
\else
The
\fi
sessions started failing due to timeouts at a packet loss rate of 10\,\%.

\message{Including measurement diagram: figures/smc/smc_executiontime_by_packetloss_paper}
\begin{figure}[t!]
\resizebox{\columnwidth}{!}{%
  \input{figures/smc/smc_executiontime_by_packetloss_paper.tikz}
}
\caption{\defaultcaption{packet loss}{\extime}}
\label{fig:smc_executiontime_by_packetloss}
\end{figure}


Comparatively weak constraints are given by the transmission rate (cf. Figure~\ref{fig:smc_executiontime_by_bandwidth}).
A very low rate of 1\,MBit does influence execution time negatively, but already between 10\,MBit and 100\,MBit all
rate-induced impediments are resolved.

Inspection shows that a transmission consists of sending a share from one host to another.
This encompasses one to maximally two packets each having only a length between 100 and 1000\,Bytes.
This is the reason why network latency has stronger influence than the transmission rate.

\message{Including measurement diagram: figures/smc/smc_executiontime_by_bandwidth_paper}
\begin{figure}[t!]
\resizebox{\columnwidth}{!}{%
\begin{tikzpicture}[x=1pt,y=1pt]
\definecolor{fillColor}{RGB}{255,255,255}
\path[use as bounding box,fill=fillColor,fill opacity=0.00] (0,0) rectangle (252.94, 93.95);
\begin{scope}
\path[clip] (  0.00,  0.00) rectangle (252.94, 93.95);
\definecolor{drawColor}{RGB}{255,255,255}
\definecolor{fillColor}{RGB}{255,255,255}

\path[draw=drawColor,line width= 0.6pt,line join=round,line cap=round,fill=fillColor] (  0.00,  0.00) rectangle (252.94, 93.95);
\end{scope}
\begin{scope}
\path[clip] ( 28.87, 27.13) rectangle (247.44, 88.45);
\definecolor{fillColor}{RGB}{255,255,255}

\path[fill=fillColor] ( 28.87, 27.13) rectangle (247.44, 88.45);
\definecolor{drawColor}{gray}{0.92}

\path[draw=drawColor,line width= 0.3pt,line join=round] ( 28.87, 29.88) --
	(247.44, 29.88);

\path[draw=drawColor,line width= 0.3pt,line join=round] ( 28.87, 53.74) --
	(247.44, 53.74);

\path[draw=drawColor,line width= 0.3pt,line join=round] ( 28.87, 77.61) --
	(247.44, 77.61);

\path[draw=drawColor,line width= 0.6pt,line join=round] ( 28.87, 41.81) --
	(247.44, 41.81);

\path[draw=drawColor,line width= 0.6pt,line join=round] ( 28.87, 65.68) --
	(247.44, 65.68);

\path[draw=drawColor,line width= 0.6pt,line join=round] ( 60.09, 27.13) --
	( 60.09, 88.45);

\path[draw=drawColor,line width= 0.6pt,line join=round] (112.14, 27.13) --
	(112.14, 88.45);

\path[draw=drawColor,line width= 0.6pt,line join=round] (164.18, 27.13) --
	(164.18, 88.45);

\path[draw=drawColor,line width= 0.6pt,line join=round] (216.22, 27.13) --
	(216.22, 88.45);
\definecolor{drawColor}{RGB}{0,0,0}

\node[text=drawColor,anchor=base,inner sep=0pt, outer sep=0pt, scale=  0.85] at ( 60.09, 70.79) {28.37};

\node[text=drawColor,anchor=base,inner sep=0pt, outer sep=0pt, scale=  0.85] at (112.14, 44.44) {7.33};

\node[text=drawColor,anchor=base,inner sep=0pt, outer sep=0pt, scale=  0.85] at (164.18, 38.91) {5.02};

\node[text=drawColor,anchor=base,inner sep=0pt, outer sep=0pt, scale=  0.85] at (216.22, 38.98) {5.05};

\path[draw=drawColor,line width= 0.6pt,line join=round] ( 60.09, 85.66) --
	(112.14, 35.45) --
	(164.18, 29.91) --
	(216.22, 29.99);
\definecolor{fillColor}{RGB}{0,0,0}

\path[draw=drawColor,line width= 0.4pt,line join=round,line cap=round,fill=fillColor] ( 60.09, 85.66) circle (  1.96);

\path[draw=drawColor,line width= 0.4pt,line join=round,line cap=round,fill=fillColor] (112.14, 35.45) circle (  1.96);

\path[draw=drawColor,line width= 0.4pt,line join=round,line cap=round,fill=fillColor] (164.18, 29.91) circle (  1.96);

\path[draw=drawColor,line width= 0.4pt,line join=round,line cap=round,fill=fillColor] (216.22, 29.99) circle (  1.96);
\definecolor{drawColor}{gray}{0.20}

\path[draw=drawColor,line width= 0.6pt,line join=round,line cap=round] ( 28.87, 27.13) rectangle (247.44, 88.45);
\end{scope}
\begin{scope}
\path[clip] (  0.00,  0.00) rectangle (252.94, 93.95);
\definecolor{drawColor}{gray}{0.30}

\node[text=drawColor,anchor=base east,inner sep=0pt, outer sep=0pt, scale=  0.56] at ( 23.92, 39.88) {10};

\node[text=drawColor,anchor=base east,inner sep=0pt, outer sep=0pt, scale=  0.56] at ( 23.92, 63.75) {20};
\end{scope}
\begin{scope}
\path[clip] (  0.00,  0.00) rectangle (252.94, 93.95);
\definecolor{drawColor}{gray}{0.20}

\path[draw=drawColor,line width= 0.6pt,line join=round] ( 26.12, 41.81) --
	( 28.87, 41.81);

\path[draw=drawColor,line width= 0.6pt,line join=round] ( 26.12, 65.68) --
	( 28.87, 65.68);
\end{scope}
\begin{scope}
\path[clip] (  0.00,  0.00) rectangle (252.94, 93.95);
\definecolor{drawColor}{gray}{0.20}

\path[draw=drawColor,line width= 0.6pt,line join=round] ( 60.09, 24.38) --
	( 60.09, 27.13);

\path[draw=drawColor,line width= 0.6pt,line join=round] (112.14, 24.38) --
	(112.14, 27.13);

\path[draw=drawColor,line width= 0.6pt,line join=round] (164.18, 24.38) --
	(164.18, 27.13);

\path[draw=drawColor,line width= 0.6pt,line join=round] (216.22, 24.38) --
	(216.22, 27.13);
\end{scope}
\begin{scope}
\path[clip] (  0.00,  0.00) rectangle (252.94, 93.95);
\definecolor{drawColor}{gray}{0.30}

\node[text=drawColor,anchor=base,inner sep=0pt, outer sep=0pt, scale=  0.56] at ( 60.09, 18.32) {1};

\node[text=drawColor,anchor=base,inner sep=0pt, outer sep=0pt, scale=  0.56] at (112.14, 18.32) {10};

\node[text=drawColor,anchor=base,inner sep=0pt, outer sep=0pt, scale=  0.56] at (164.18, 18.32) {100};

\node[text=drawColor,anchor=base,inner sep=0pt, outer sep=0pt, scale=  0.56] at (216.22, 18.32) {1000};
\end{scope}
\begin{scope}
\path[clip] (  0.00,  0.00) rectangle (252.94, 93.95);
\definecolor{drawColor}{RGB}{0,0,0}

\node[text=drawColor,anchor=base,inner sep=0pt, outer sep=0pt, scale=  0.70] at (138.16,  8.00) {Transmission rate [Mbit]};
\end{scope}
\begin{scope}
\path[clip] (  0.00,  0.00) rectangle (252.94, 93.95);
\definecolor{drawColor}{RGB}{0,0,0}

\node[text=drawColor,rotate= 90.00,anchor=base,inner sep=0pt, outer sep=0pt, scale=  0.70] at ( 10.32, 57.79) {Time [s]};
\end{scope}
\end{tikzpicture}
}
\caption{\defaultcaption{transmission rate}{\extime}}
\label{fig:smc_executiontime_by_bandwidth}
\end{figure}

%

In conclusion, each single computation has a low duration; the overall duration increases linearly with the number of peers.
While this influence is comparatively small, the network parameters have \iflongversion -- as expected -- \fi the highest influence on the execution time.
In the ranges of the practically relevant intervals \iflongversion chosen for these parameters during the tests,\fi we saw that the transmission rate can influence the execution time by factor 5, packet loss has an influence up to a factor of approximately 110 and network latency can slow down the computation even by factor 550.
These impediments already occur at network configurations which are realistic on the Internet or on the mobile Internet at least.

\subsection{Parallelization}
\label{sec:paral}
\renewcommand{\reason}{parallelization}
\label{sec:paral_invocations}
\iflongversion
In Section \ref{sec:performance_model} we modeled the execution as alternating sequence of computation and communication steps.
This characteristic leads to the existence of two potential bottlenecks:
\else
Two potential bottlenecks can exist (cf. Section~\ref{sec:performance_model}):
\fi
The computation steps are constrained by the host and the communication steps by the links between them. Due to their strict sequential execution, the occurring delays are additive.

This situation can be generically improved in certain scenarios:
In our test setup we performed 1000 calculations sequentially.
When a given number of input data is known in advance, multiple computations can be performed at the same time.
\FRESCO provides a \texttt{ParallelProtocolProducer} which allows combining subprotocols so that they are executed in parallel using individual threads.
\iflongversion

We used this feature in order to determine, whether parallelization yields a benefit regarding execution time and how it influences the remaining resource utilization of \FRESCO.

\fi
Parallelization was parametrized with the values $\{1,5,10,20,50,100,200,500,1000\}$.
\iflongversion Higher levels of parallelization are not possible in our setup as the overall input per node was also constrained to 1000 data points.\fi

Our initial finding is that the benefit of parallelization depends on the size of the delay introduced by host or network parameters.
Figure \ref{fig:smc_executiontime_by_paral_zero}  shows that in our default setting parallelization reduces an initial duration of 4.910 seconds with \textit{pf} $== 1 $ in average to 3.631 with a \textit{pf} $== 500$, a reduction by 26\,\%.
This improvement depends on the availability of multiple CPU cores.
Otherwise, parallelization leads to a slight rise in execution time\iflongversion\else{} due to its organizational overhead\fi.

\message{Including measurement diagram: figures/smc/smc_executiontime_by_paral_detail_netlat_0_paper}
\begin{figure}[t!]
\resizebox{\columnwidth}{!}{%
  \input{figures/smc/smc_executiontime_by_paral_detail_netlat_0_paper.tikz}
}
\caption{Impact of network latency on the execution time depending on parallelization with 0\,ms additional latency}
\label{fig:smc_executiontime_by_paral_zero}
\end{figure}


However, when examining cases with e.g. higher network latency (Figure~\ref{fig:smc_executiontime_by_paral_fivehundred}), a higher reduction can be achieved.
An initial duration of 2315 seconds is reduced to 435 seconds when \textit{pf} $== 200$, a reduction by
81\,\%.

\message{Including measurement diagram: figures/smc/smc_executiontime_by_paral_detail_netlat_500_paper}
\begin{figure}[t!]
\resizebox{\columnwidth}{!}{%
  \input{figures/smc/smc_executiontime_by_paral_detail_netlat_500_paper.tikz}
}
\caption{Impact of network latency on the execution time depending on parallelization with 500\,ms additional latency}
\label{fig:smc_executiontime_by_paral_fivehundred}
\end{figure}

\iflongversion
We make two observations:
Firstly, parallelization imposes a computational overhead itself.
This becomes visible when examining cases where a single core is
utilized as well as the cases where a very small parallelization factor
(\textit{pf} $== 5$) is chosen.
Then, the overall duration increases.
Secondly, the optimal value for \textit{pf} is not fixed but depends on aspects as the specific protocol to be executed as well as the network and host properties likewise.

Nevertheless, it is obvious that parallelization of former sequential computations can yield a significant benefit with respect to  execution time.
\fi
Investigation of the code shows that the network layer does not combine messages of different but simultaneously performed sessions; they are sent via different ``channels''.
Instead, this layer processes the messages to be sent in a strict sequential fashion.
This means that the measured improvements result from the time-multiplex utilization of the networking layer:
During sequential execution of subsequent computations, all other parties often wait for a single party to perform its computation and sending of shares.
With parallelization, we achieve that every participant is
kept busy on the communication part -- the bottleneck -- all the time.


When employing parallelization on a basic TTP solution a reduction of the \emph{actual communication overhead} is possible.
Input data of multiple computations can be combined into a single (or a low amount of) packets.
In consequence, the impact of network latency can be  highly reduced.\footnote{In Figure~\ref{fig:smc_executiontime_by_paral_fivehundred} we show an estimation of the TTP performance while neglecting that a high parallelization rate makes it necessary to split the packets again for transmission.}


%


\message{Including measurement diagram: figures/smc/smc_cycles_by_netlat_paral_paper}
\begin{figure}[t!]
\resizebox{\columnwidth}{!}{%
  \input{figures/smc/smc_cycles_by_netlat_paral_paper.tikz}
}
\caption{\defaultcaption{\reason}{\cpucycles}}
\label{fig:smc_cycles_by_netlat_paral}
\end{figure}

Figure \ref{fig:smc_cycles_by_netlat_paral} shows the consumed CPU cycles depending on the degree of parallelization with non-zero additional network latency and 8 available cores per host.
Parallelization does not impose notable penalties on the hosts'
\iflongversion
CPU.
The amount of consumed cycles stays roughly the same as the amount of computations is only increased by a small computational overhead.
\else
CPU.
\fi
Correspondingly, as the execution duration is significantly reduced, the actual degree of CPU utilization during that time increases.

Regarding transferred data, an increasing degree of parallelization does not lead to a corresponding increase (cf. Figure \ref{fig:smc_bytes_by_netlat_paral}).
Instead, it is still bounded by approximately 5.7\,MBytes.



\message{Including measurement diagram: figures/smc/smc_packets_by_netlat_paral_paper}
\begin{figure}[t!]
\resizebox{\columnwidth}{!}{%
  \input{figures/smc/smc_packets_by_netlat_paral_paper.tikz}
}
\caption{\defaultcaption{\reason}{\packets}}
\label{fig:smc_bytes_by_netlat_paral}
\end{figure}

\ifincludestackmem
\iflongversion
When regarding the maximum stack memory with no parallelization and changing network latency (Figure \ref{fig:smc_max_memory_stack_by_netlat}), the memory consumption of cases with no additional latency added is relatively higher compared to all other cases. This vanishes (Figure \ref{fig:smc_max_memory_stack_by_netlat_paral}) with an increasing parallelization factor; beginning at \textit{pf} \(== 10 \), the consumption is as high as with additional latency.
However,  in general
\else
Figure \ref{fig:smc_max_memory_stack_by_netlat_paral} shows that
\fi
the stack memory is highly influenced by the heuristics of the garbage collector, causing indistinct trends and high variation.
Nevertheless, these variations stay within a small range.
I.e.  parallelization does not have any negative impact on the stack memory utilization.

\message{Including measurement diagram: figures/smc/smc_max_memory_stack_by_netlat_paral_paper}
\begin{figure}[t!]
\resizebox{\columnwidth}{!}{%
  \input{figures/smc/smc_max_memory_stack_by_netlat_paral_paper.tikz}
}
\caption{\defaultcaption{\reason}{\stackmem}}
\label{fig:smc_max_memory_stack_by_netlat_paral}
\end{figure}

\fi

Figure \ref{fig:smc_max_memory_heap_by_netlat_paral} shows that parallelization leads to a late and not completely clear trend of memory increase when \textit{pf} $ \ge 50$.
This could extend to higher degrees of parallelization.
However, the benefits of parallelization are achieved with a much lower
\textit{pf} than where memory consumption starts to increase gradually.
That means a sweet spot of beneficial parallelization without memory penalties should be generally identifiable.
Nevertheless this is a trend where further investigations could provide more insights on whether memory consumption is bounded or not.

\iflongversion
While this is the general tendency, some details are notable:
Firstly, the memory consumption of computations with no additional network latency are less consistent and have a wider variation.
Secondly, with an additional network latency of 8 or 25 milliseconds, the memory consumption seems to decrease around a parallelization of 50 to 500 points per computation. Only at 1000 points, they follow the general trend of increase at the same height.

Having a standard value of 70\,MBytes is in line with the other experiments.
The first effect of variation can be explained as follows: Having no additional network latency which dominates the natural latency, the natural fluctuations might influence the garbage collector (which is the main influencing factor of maximum memory usage) more and in other ways.
We also explain the second effect similarly: While the garbage collector's behavior can hardly be predicted, it is possible that these combinations of a parallelization factor and the network latency influences it to be invoked slightly earlier.
\fi

\message{Including measurement diagram: figures/smc/smc_max_memory_heap_by_netlat_paral_paper}
\begin{figure}[t!]
\resizebox{\columnwidth}{!}{%
  \input{figures/smc/smc_max_memory_heap_by_netlat_paral_paper.tikz}
}
\caption{\defaultcaption{\reason}{\heapmem}}
\label{fig:smc_max_memory_heap_by_netlat_paral}
\end{figure}

In conclusion, we could show that parallelization is able to reduce the computation duration approximately by factor 5.
Therefor, computation of 20 items in parallel is sufficient and already exploits the full parallelization potential.
Further increase of the parallelization factor did not yield notable improvements.
However, different types of computations yield protocols which differ in their ability to be parallelized.
Hence, the parallelization factor should be newly determined every time a new protocol is used.
Finally, while parallelization yields a notable improvement on execution time, it does not cause any changes regarding the CPU cycles.
This shows that parallelization can effectively be used to utilize all available CPU resources.
Concomitantly, the former sequential execution, creating two alternating bottlenecks -- the CPU and the network -- is changed to a parallel execution of
both phases (with respect to different computations). In consequence, only the stricter of both bottlenecks constitutes the limiting factor instead of their sum.
In any case, multi-core hosts are necessary in order to leverage parallelization advantages at all.

%

\section{Practical Implications}
\label{sec:practical_implications}
Our results show that \FRESCO as an implementation of SMC possesses a performance and resource utilization behavior which allows practical application:
In the setting of an intranet, computations are efficiently performed.
The execution time is around 2 to 3\,ms per session and peer.
This allows batch processing of data and interactive use cases.
Performance might, however, not be sufficient for the realization of real-time applications depending on the computation to be carried out.
Regarding the hosts systems, multiple cores are necessary when parallelization can be  utilized. In other cases, secure computation should also be feasible with weaker devices.
Memory consumption can become critical when a multitude of peers participates in the computations.
This must be considered upon productive use.
However, regarding all identified performance results, we deem the memory consumption to be more related to Java than to secret sharing or SMC in general. Having a setting of memory constrained devices, a more economical programming language would be more appropriate.

In wide area networks as the Internet and possibly mobile Internet, network latency is the most influential constraining factor.
Execution time degrades strongly with increasing latency.
In these contexts, we currently only see batch processing as a use case:
Given it is acceptable to wait several minutes for a computation result, SMC can be utilized.
However, in this context it is more likely that parallelization can be applied, which
decreases the latency penalties to some degree.
Further improvement of the situation would require to reduce the amount of transmitted packets.
This could be possible by stricter orchestration of computations running in parallel, where packets between different peers would be used for multiple sessions simultaneously.
\iflongversion
Given each participant has k input values available and all protocols are executed completely simultaneously and synchronized.
Then, k shares from the same source to the same destination could possibly combined to a single message, effectively reducing the communication duration by factor k.
\fi
On contrary, our current solution applies parallelization which does not enforce message combination, but only enabled waiting times per host to be used for further computations.

\section{Conclusion}
\label{sec:conclusion}

This paper presents the results of thorough measurements to assess the fundamental practical applicability of secure multiparty computation (SMC) in real-world contexts.

We show how SMC sessions can be understood with regard to performance as a alternating sequence of local computation and communication between participating peers. This yields practical implications, bearing in mind that both types have their own individual bottleneck: Typically, their delay is strictly summative during a single execution.

In our measurements, we
examine how network latency, transmission rate and packet loss, as  well as the number of peers influence the execution time, the CPU utilization, memory allocation and the amount of transmitted data.
Furthermore, we analyze whether parallelization of formerly consecutive sessions can overcome the additivity of delays.

Interpreting our findings, we conclude that SMC is practically applicable with weak limitations in intranet settings. Here, requirements for participating host systems are in ranges of today's commodity hardware.
Furthermore, SMC seems to be applicable to some (lesser) degree in Internet settings. Here, network latency has the biggest negative influence on performance.
However, as performance of SMC protocols continues to increase, we expect that feasibility of SMC over the Internet will also improve in the next years.

%
%
\section{Acknowledgements}
We would like to thank Daniel Raumer and Florian Wohlfart for their valuable feedback on the initial versions of the paper.

\bibliographystyle{IEEEtran}
\bibliography{literature}

\begin{thebibliography}{10}
\providecommand{\url}[1]{#1}
\csname url@samestyle\endcsname
\providecommand{\newblock}{\relax}
\providecommand{\bibinfo}[2]{#2}
\providecommand{\BIBentrySTDinterwordspacing}{\spaceskip=0pt\relax}
\providecommand{\BIBentryALTinterwordstretchfactor}{4}
\providecommand{\BIBentryALTinterwordspacing}{\spaceskip=\fontdimen2\font plus
\BIBentryALTinterwordstretchfactor\fontdimen3\font minus
  \fontdimen4\font\relax}
\providecommand{\BIBforeignlanguage}[2]{{%
\expandafter\ifx\csname l@#1\endcsname\relax
\typeout{** WARNING: IEEEtran.bst: No hyphenation pattern has been}%
\typeout{** loaded for the language `#1'. Using the pattern for}%
\typeout{** the default language instead.}%
\else
\language=\csname l@#1\endcsname
\fi
#2}}
\providecommand{\BIBdecl}{\relax}
\BIBdecl

\bibitem{Yao1982}
A.~C. Yao, ``{Protocols for secure computations},'' in \emph{Proceedings of the
  23rd Annual Symposium on Foundations of Computer Science}.\hskip 1em plus
  0.5em minus 0.4em\relax Washington, DC, USA: IEEE, 1982, pp. 1--5.

\bibitem{SugarBeetSMC}
P.~Bogetoft, D.~L. Christensen, I.~Damg{\aa}rd, M.~Geisler, T.~Jakobsen,
  M.~Kr{\o}igaard, J.~D. Nielsen, J.~B. Nielsen, K.~Nielsen, J.~Pagter,
  M.~Schwartzbach, and T.~Toft, ``{Secure multiparty computation goes live},''
  in \emph{Lecture Notes in Computer Science}, vol. 5628 LNCS, 2009, pp.
  325--343.

\bibitem{SEPIANetwork}
M.~Burkhart, M.~Strasser, D.~Many, and X.~Dimitropoulos, ``{SEPIA:
  Privacy-preserving Aggregation of Multi-domain Network Events and
  Statistics},'' \emph{Proceedings of the 19th USENIX Conference on Security},
  p.~15, 2010.

\bibitem{SMCAviation}
M.~Zanin, T.~T. Delibasi, J.~C. Triana, V.~Mirchandani, E.~{{\'{A}}lvarez
  Pereira}, A.~Enrich, D.~Perez, C.~Paşaoğlu, M.~Fidanoglu, E.~Koyuncu,
  G.~Guner, I.~Ozkol, and G.~Inalhan, ``{Towards a secure trading of aviation
  CO2 allowance},'' \emph{Journal of Air Transport Management}, vol.~56, pp.
  3--11, 2016.

\bibitem{SMCFinancial}
D.~Bogdanov, R.~Talviste, and J.~Willemson, ``{Deploying secure multi-party
  computation for financial data analysis},'' \emph{Financial Cryptography},
  pp. 57 -- 64, 2012.

\bibitem{Yao1986}
A.~C. Yao, ``{How to generate and exchange secrets},'' in \emph{Proceedings of
  the 27th IEEE Symposium on Foundations of Computer Science}.\hskip 1em plus
  0.5em minus 0.4em\relax IEEE Computer Society Press, 1986, pp. 162--167.

\bibitem{Rivest1978a}
R.~L. Rivest, L.~Adleman, and M.~L. Dertouzos, ``{On Data Banks and Privacy
  Homomorphisms},'' \emph{Foundations of Secure Computation}, pp. 169--180,
  1978.

\bibitem{Shamir1979}
A.~Shamir, ``{How To Share a Secret},'' \emph{Communications of the ACM
  (CACM)}, vol.~22, no.~11, pp. 612--613, 1979.

\bibitem{BGW88}
M.~Ben-Or, S.~Goldwasser, and A.~Wigderson, ``{Completeness Theorems for
  Non-Cryptographic Fault Tolerant Distributed Computation},''
  \emph{Proceedings of the 20th Annual ACM Symposium on the Theory of Computing
  (STOC)}, pp. 1--10, 1988.

\bibitem{Goldreich1987}
O.~Goldreich, S.~Micali, and A.~Wigderson, ``{How to play ANY mental game},''
  in \emph{Proceedings of the Nineteenth Annual ACM Conference on Theory of
  Computing -- STOC '87}.\hskip 1em plus 0.5em minus 0.4em\relax New York, NY,
  USA: ACM, 1987, pp. 218--229.

\bibitem{bmr}
D.~Beaver, S.~Micali, and P.~Rogaway, ``{The Round Complexity of Secure
  Protocols},'' in \emph{Proceedins of the 22nd Annual ACM Symposium on the
  Theory of Computing}, 1990, pp. 503----513.

\bibitem{Chaum1988}
D.~Chaum, I.~B. Damg{\aa}rd, and J.~van~de Graaf, ``{Multiparty Computations
  Ensuring Privacy of Each Party's Input and Correctness of the Result},'' in
  \emph{Advances in Cryptology}, C.~Pomerance, Ed.\hskip 1em plus 0.5em minus
  0.4em\relax Berlin Heidelberg: Springer-Verlag, 1987, pp. 87--119.

\bibitem{Chaum1988a}
D.~Chaum, C.~Cr{\'{e}}peau, and I.~Damg{\aa}rd, ``{Multiparty Unconditionally
  Secure Protocols},'' \emph{Proceedings of the 20th Annual ACM Symposium on
  Theory of Computing}, pp. 11--19, 1988.

\bibitem{Rabin1989}
T.~Rabin and M.~Ben-Or, ``{Verifiable secret sharing and multiparty protocols
  with honest majority},'' \emph{Proceedings of the 21st Annual ACM Symposium
  on Theory of Computing}, pp. 73--85, 1989.

\bibitem{VIFF}
M.~Geisler, ``{Cryptographic Protocols: Theory and Implementation},'' Ph.D.
  dissertation, Aarhus University, 2010.

\bibitem{FairplayMP}
A.~Ben-David, N.~Nisan, and B.~Pinkas, ``{FairplayMP: A System for Secure
  Multi-Party Computation},'' \emph{Proceedings of the 15th ACM Conference on
  Computer and Communications Security}, pp. 257--266, 2008.

\bibitem{Sharemind}
D.~Bogdanov, S.~Laur, and J.~Willemson, ``{Sharemind: A framework for fast
  privacy-preserving computations},'' in \emph{IACR Cryptology ePrint
  Archive}.\hskip 1em plus 0.5em minus 0.4em\relax Springer, 2008, no. October,
  p. 289.

\bibitem{BristolMPC}
I.~Damg{\aa}rd, V.~Pastro, N.~Smart, and S.~Zakarias, ``{Multiparty computation
  from somewhat homomorphic encryption},'' in \emph{Lecture Notes in Computer
  Science}, vol. 7417, 2012, pp. 643--662.

\bibitem{MASCOT}
M.~Keller, E.~Orsini, and P.~Scholl, ``{MASCOT},'' in \emph{Proceedings of the
  2016 ACM SIGSAC Conference on Computer and Communications Security}, 2016,
  pp. 830--842.

\bibitem{frescoURL}
\BIBentryALTinterwordspacing
``{A FRamework for Efficient Secure COmputation}.'' [Online]. Available:
  \url{https://github.com/aicis/fresco}
\BIBentrySTDinterwordspacing

\bibitem{SharemindPhD}
D.~Bogdanov, ``{Sharemind: programmable secure computations with practical
  applications},'' Ph.D. dissertation, 2013.

\bibitem{Cybernetica}
\BIBentryALTinterwordspacing
``{Cybernetica},'' 2017. [Online]. Available: \url{https://www.cyber.ee}
\BIBentrySTDinterwordspacing

\bibitem{Damgard2017a}
I.~Damg{\aa}rd, K.~Damg{\aa}rd, K.~Nielsen, P.~S. Nordholt, and T.~Toft,
  ``{Confidential benchmarking based on multiparty computation},''
  \emph{Lecture Notes in Computer Science}, vol. 9603 LNCS, pp. 169--187, 2017.

\bibitem{Thoma2012}
C.~Thoma, T.~Cui, and F.~Franchetti, ``{Secure Multiparty Computation Based
  Privacy Preserving Smart Metering System},'' \emph{44th North American Power
  Symposium (NAPS)}, pp. 1--6, 2012.

\bibitem{Bogetoft2006}
P.~Bogetoft, I.~Damg{\aa}rd, T.~Jakobsen, K.~Nielsen, J.~Pagter, and T.~Toft,
  ``{A Practical Implementation of Secure Auctions Based on Multiparty Integer
  Computation},'' \emph{Financial Cryptography}, vol. 4107, pp. 142--147, 2006.

\bibitem{Bonawitz2017}
K.~Bonawitz, V.~Ivanov, B.~Kreuter, A.~Marcedone, H.~B. McMahan, S.~Patel,
  D.~Ramage, A.~Segal, and K.~Seth, ``{Practical Secure Aggregation for Privacy
  Preserving Machine Learning.}'' \emph{IACR Cryptology ePrint Archive}, vol.
  2017, p. 281, 2017.

\bibitem{practicaltwoparty}
B.~Pinkas, T.~Schneider, N.~P. Smart, and S.~C. Williams, ``{Secure two-party
  computation is practical},'' \emph{Lecture Notes in Computer Science}, vol.
  5912 LNCS, pp. 250--267, 2009.

\bibitem{hpsmc}
D.~Bogdanov, M.~Niitsoo, T.~Toft, and J.~Willemson, ``{High-performance secure
  multi-party computation for data mining applications},'' \emph{International
  Journal of Information Security}, vol.~11, no.~6, pp. 403--418, 2012.

\bibitem{comparisonperformance}
F.~Kerschbaum, D.~Biswas, and S.~{De Hoogh}, ``{Performance comparison of
  secure comparison protocols},'' \emph{Proceedings - International Workshop on
  Database and Expert Systems Applications, DEXA}, no. October, pp. 133--136,
  2009.

\bibitem{smccommunicationcomplexity}
F.~Kerschbaum, D.~Dahlmeier, A.~Schr{\"{o}}pfer, and D.~Biswas, ``{On the
  practical importance of communication complexity for secure multi-party
  computation protocols},'' \emph{Proceedings of the 2009 ACM symposium on
  Applied Computing - SAC '09}, pp. 2008--2015, 2009.

\bibitem{Canetti1999}
R.~Canetti, ``{Security and Composition of Multi-party Cryptographic
  Protocols},'' 1999.

\bibitem{measrdroid}
\BIBentryALTinterwordspacing
{Chair of Network Architectures and Services; TUM}, ``{MeasrDroid}.'' [Online].
  Available: \url{http://www.droid.net.in.tum.de}
\BIBentrySTDinterwordspacing

\bibitem{raspi}
\BIBentryALTinterwordspacing
``{Raspberry Pi Models}.'' [Online]. Available:
  \url{https://www.raspberrypi.org/products/}
\BIBentrySTDinterwordspacing

\end{thebibliography}

\end{document}